# Effect of crystallographic dislocations on the reverse performance of 4H-SiC p–n diodes

Feng Zhao *, Mohammad M. Islam, Biplob K. Daas, Tangali S. Sudarshan

Q1 *Department of Electrical Engineering, University of South Carolina, Columbia, SC 29208, United States*

ABSTRACT

A quantitative study was performed to investigate the impact of crystallographic dislocation defects, including screw dislocation, basal plane dislocation, and threading edge dislocation, and their locations in active and JTE region, on the reverse performance of 4H-SiC p–n diodes. It was found that higher leakage current in diodes is associated with basal plane dislocations, while lower breakdown voltage is attributed to screw dislocations. The above influence increases in severity when the dislocation is in the active region than in the JTE region. Furthermore, due to the closed-core nature, the impact of threading edge dislocation on the reverse performance of the p–n diodes is less severe than that of other dislocations although its density is much higher.

## 1. Introduction

With the need for new generation compact and low-loss high power electronics capable of operating at extreme conditions, SiC based devices are being investigated extensively due to its outstanding material properties [1]. However, a significant roadblock for the development and commercialization of high power SiC devices is due to the unavailability of high quality epitaxial substrates with low density of defects.

The high density of crystallographic defects has been a long-standing problem with SiC material. Micropipes are known to significantly affect device characteristics such as leakage current and breakdown voltage in p–n and Schottky diodes [2,3]. Due to improvements in crystal-growth technology during the past several years, micropipe density has been significantly reduced and 4-inch Zero-Micropipe n-type SiC substrates are currently available [4]. Even though not as detrimental to device performance as micropipes, the adverse impact of other crystallographic defects, especially the screw dislocation, basal plane dislocation, and threading edge dislocation on the device performance needs to be investigated. Qualitative studies show that 4H-SiC p–n diodes and Schottky diodes containing screw dislocations exhibit a high reverse leakage current and low reverse breakdown voltage [5,6]. Basal plane dislocations and their influence on the forward I–V of p–n diodes and BJTs were also investigated extensively [7,8]. Threading edge dislocation pairs also compromise the performance of SiC devices and the role of such defects on the reverse leakage currents of 4H-SiC junction barrier Schottky (JBS) diodes were reported [6].

The focus of this work is to correlate the crystallographic dislocations with the reverse performance of 4H-SiC p–n diodes, and quantify the impact of screw dislocation, basal plane dislocation, and threading edge dislocation as well as their locations in active and JTE regions on breakdown voltage and reverse leakage current. This study sets directions for essential improvements in SiC material quality that will result in higher yield of commercial devices.

## 2. Experiment

The $n^-$ epitaxial layer with a thickness of 15 μm and nitrogen concentration of $5\times10^{15}\,cm^{-3}$ was grown on a commercial 8° off-axis n-type 4H-SiC substrate in a hot-wall CVD reactor with precursor's silane (3 sccm) and propane (0.9 sccm) with hydrogen as carrier gas at 1600 °C and 300 torr. The wafer was then diced into small pieces with an area of $8\times8\,mm^2$ for diode fabrication. The epi-structure and edge termination were designed for 2500 V blocking voltage based on simulation by ATLAS software of Silvaco. The p-type active region with an area of $1\times1\,mm^2$ and the JTE region with a width of 150 μm were formed simultaneously by an aluminum implantation at room temperature followed by an annealing process at 1500 °C for 30 min in argon ambient to activate the dopant. A 1 μm-thick graphite cap layer was used to protect the sample from surface degradation due to Si evaporation from SiC during annealing. Standard lithography was performed, and Ti/Al/Ti/Ni metal stack for p-type anode contact was deposited by e-beam evaporation and a lift-off process. Ni was deposited on the back of the sample for n-type cathode contact. Both n- and p-type ohmic contacts were prepared by a rapid thermal annealing (RTA) process at 1000 °C for 1 min in high-purity nitrogen gas.

* Corresponding author. Tel.: +1 803 777 6303; fax: +1 803 777 8045.
*E-mail address:* zhaof2@cec.sc.edu (F. Zhao).

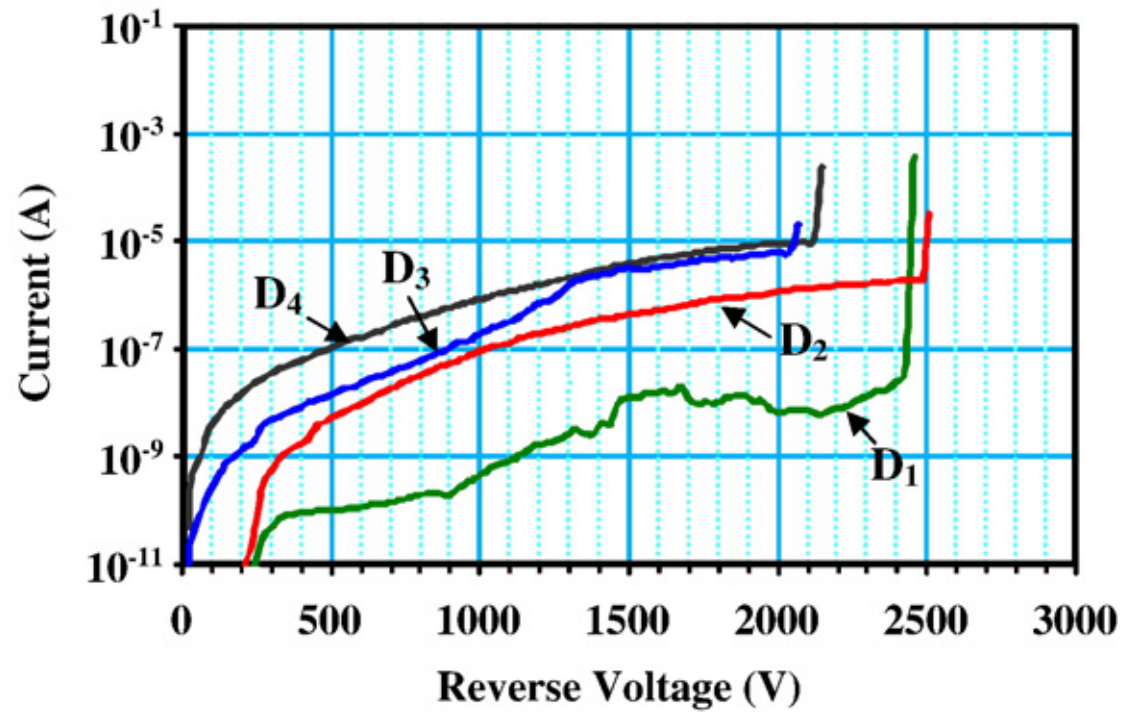

**Fig. 1.** The reverse I–V characteristics of 4 identical p–n diodes ($D_1$–$D_4$) from different portions on the 4H-SiC sample.

**Table 1**
Threshold breakdown voltage, leakage current, and the number and location of dislocation defects in 4 identical diodes under study.

| Diode # | $V_{BR}$(V) @10 µA | $I_L$(µA) @2000 V | In active region | | | In JTE region | | |
|---|---|---|---|---|---|---|---|---|
| | | | $n_{SD}$ | $n_{BPD}$ | $n_{TED}$ | $n_{SD}$ | $n_{BPD}$ | $n_{TED}$ |
| $D_1$ | 2460 | 0.007 | 1 | 0 | 110 | 0 | 2 | 47 |
| $D_2$ | 2510 | 1.13 | 0 | 0 | 76 | 0 | 7 | 41 |
| $D_3$ | 2070 | 6.3 | 5 | 2 | 85 | 0 | 2 | 55 |
| $D_4$ | 2150 | 8.6 | 3 | 2 | 67 | 2 | 7 | 35 |

Reverse current–voltage characterization was performed using a Keithley 2410 voltage source measurement unit in low light condition with the current compliance set to 1 mA (0.01 A/cm$^2$). Fluorinert was used during high voltage testing over 1000 V to mitigate surface flash-over. After current–voltage testing, metal contacts were removed and chemical etching in molten KOH at 600 °C was performed for 3 min to expose defects present in the sample. A Nomarsky optical microscope was used to examine the surface of the diodes to identify the type of defects by etch pit shape recognition and to count the number of defects present in both the active and JTE region of each diode. Device performance was then correlated with defects present in the diodes.

## 3. Results and discussion

Fig. 1 shows the reverse I–V characteristics on a logarithmic scale of 4 identical p–n diodes ($D_1$–$D_4$) from different portions on the SiC sample. Diodes D1 and D2 clearly exhibit higher breakdown voltage and lower leakage current than diodes D3 and D4. The leakage current $I_L$ at the reverse bias of 2000 V and the threshold breakdown voltage $V_{BR}$ corresponding to the leakage current of 10 µA ($1\times10^{-3}$ A/cm$^2$) from these 4 diodes were summarized in Table 1. The variation of leakage current and breakdown voltage are directly correlated to the number of crystallographic dislocation defects and their locations in the active region and JTE region based on a quantitative study.

An optical microscopy image of SiC surface after molten KOH etching is shown in Fig. 2. Inspection of the wafer surface after chemical etching allows identification of the type of dislocations by the characteristics of the etch pits. The hexagonal etch pits correspond to screw dislocations (hollow core) and threading edge dislocations (closed core), while the oval-shaped etch pits correspond to basal plane dislocations. The number of screw dislocations, threading edge dislocations and basal plane dislocations in both the active and JTE regions on each diode was counted carefully and summarized in Table 1, along with the threshold breakdown voltage $V_{BR}$ and leakage current $I_L$.

In diodes $D_1$ and $D_2$, there is no basal plane dislocation in the active region and screw dislocation in the JTE region. The breakdown voltage $V_{BR}$ (2460 V) of $D_1$ is lower than $V_{BR}$ (2510 V) of $D_2$ due to the screw dislocation existing in its active region. When comparing the number of screw dislocations in the active region of these 4 diodes, $n_{SD2} < n_{SD1} < n_{SD4} < n_{SD3}$, it clearly shows that more screw dislocations result in lower breakdown voltage $V_{BR2} > V_{BR1} > V_{BR4} > V_{BR3}$. There are 7 basal plane dislocations in the JTE region of diode $D_2$, and 2 in the JTE region of diode $D_1$. Since $V_{BR2} > V_{BR1}$, basal plane dislocations have less influence on breakdown voltage, but more basal plane dislocations lead to a higher leakage current of 1.13 μA in diode $D_2$ than 0.007 μA in diode $D_1$. There are 5 screw dislocations found in the active region of diode $D_3$, resulting in a much lower breakdown voltage (2070 V) compared to those of diode $D_1$ and $D_2$. The leakage current (6.3 μA) of $D_3$ is also higher than those of diode $D_1$ and $D_2$ due to the basal plane dislocations in its active region, although the number of basal plane dislocation in the JTE region of diode $D_3$ is 2, not more than those in the JTE region of diode $D_1$ (2 BPDs) and $D_2$ (7 BPDs). Therefore, we conclude that when basal plane dislocations are in the active region, they affect leakage current more severely than when they exist in the JTE region. Compared to diode $D_3$ without any screw dislocation found in the JTE region, there are 2 screw dislocations in the JTE region of diode $D_4$. However, the breakdown voltage of diode $D_4$ is still higher than that of diode $D_3$, $V_{BR4} > V_{BR3}$, since there are 3 screw dislocations in the active region of diode $D_4$, less than those 5 screw dislocations in the active region of diode $D_3$. This result shows that when screw dislocations are in the active region, they affect breakdown voltage more than when they exist in the JTE region. The number of basal plane dislocations in the active region of diode $D_3$ and $D_4$ is same, while the leakage current of diode $D_4$ is higher than that of diode $D_3$ due to more basal plane dislocations in its JTE region, 7 versus 2. The density of threading edge dislocation is much higher than that of screw dislocation and basal plane dislocation, as shown in Table 1, but due to its closed-core nature, its impact on device reverse performance is much less in severity compared to the other two types of dislocations. It was reported before that threading edge dislocation pair array affects the blocking capabilities of JBS diodes [6]. To truly investigate the impact of an individual threading edge dislocation on the diode performance, diodes with small areas to only enclose threading edge dislocation need to be fabricated and characterized.

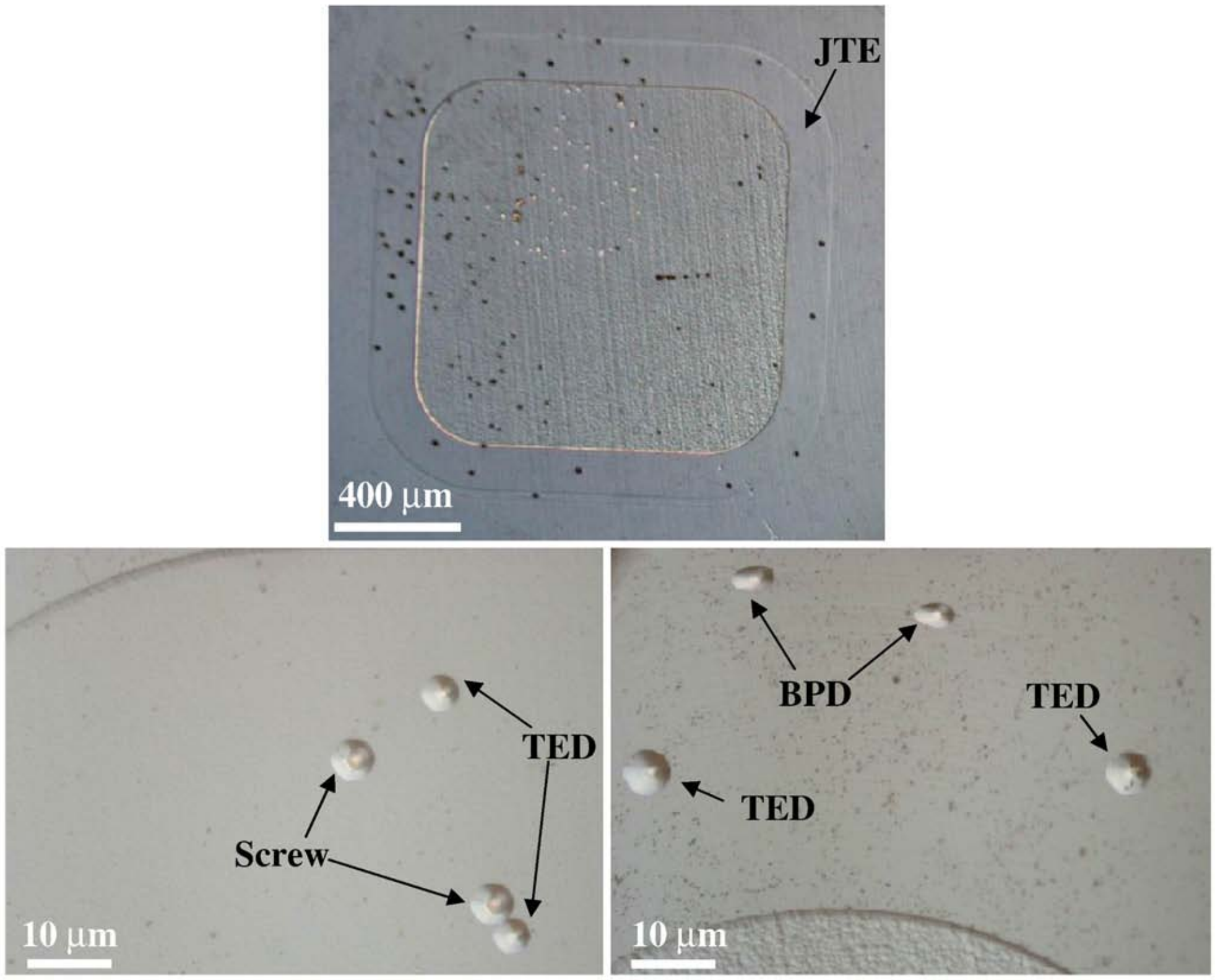

**Fig. 2.** Optical microscopy image of a 4H-SiC p–n diode after molten KOH etch showing screw dislocation (SD), basal plane dislocation (BPD), and threading edge dislocation (TED).

## 4. Conclusion

The impact of screw dislocation, basal plane dislocation, and threading edge dislocation, and their locations in the active and JTE regions, on the reverse performance of 4H-SiC p–n diodes was investigated by a quantitative study. Based on the results and analysis, we conclude that: (1) screw dislocations affect breakdown voltage while basal plane dislocations affect leakage current; (2) basal plane dislocations in the active region affect leakage current more than when they are present in the JTE region; (3) screw dislocations in the active region affect breakdown voltage more than when they are present in the JTE region; and (4) the effect of threading edge dislocation is much less severe than that of screw dislocation and basal plane dislocation.